\documentclass[english]{article}
\usepackage[latin9]{inputenc}
\usepackage{color}
\usepackage{amssymb}
\usepackage{babel}

\begin{document}

\title{\textbf{Effective temperature, Hawking radiation and quasinormal
modes}}

\author{\textbf{Christian Corda}}

\maketitle
\begin{center}
Institute for Theoretical Physics and Advanced Mathematics Einstein-Galilei,
Via Santa Gonda 14, 59100 Prato, Italy 
\par\end{center}

\begin{center}
and 
\par\end{center}

\begin{center}
Inter-University Centre Engineering of life and Environment, LIUM
University, Via Lugano 2, 6500 Bellinzona, Switzerland
\par\end{center}

\begin{center}
\textit{E-mail addresses:} \textcolor{blue}{cordac.galilei@gmail.com} 
\par\end{center}
\begin{abstract}
Parikh and Wilczek have shown that Hawking radiation's spectrum cannot
be strictly thermal. Such a non-strictly thermal character implies
that the spectrum is also not strictly continuous and thus generates
a natural correspondence between Hawking radiation and black hole's
quasinormal modes. This issue endorses the idea that, in an underlying
unitary quantum gravity theory, black holes result highly excited
states. 

We use this key point to re-analyze the spectrum of black hole's quasinormal
modes by introducing a black hole's \emph{effective temperature.}
Our analysis changes the physical understanding of such a spectrum
and enables a re-examination of various results in the literature
which realizes important modifies on quantum physics of black holes.
In particular, the formula of the horizon's area quantization and
the number of quanta of area are modified becoming functions of the
quantum {}``overtone'' number $n$. Consequently, Bekenstein-Hawking
entropy, its sub-leading corrections and the number of microstates,
i.e. quantities which are fundamental to realize unitary quantum gravity
theory, are also modified. They become functions of the quantum overtone
number too. 

Previous results in the literature are re-obtained in the very large
$n$ limit. 
\end{abstract}
\begin{center}
\emph{Essay written for the Gravity Research Foundation 2012 Awards
for Essays on Gravitation. It was awarded with an honorable mention.}
\par\end{center}

\noindent Analyzing Hawking Radiation \cite{key-1} as tunneling,
Parikh and Wilczek showed that the radiation spectrum cannot be strictly
thermal \cite{key-2,key-3}. Parikh released an intriguing physical
interpretation of this fundamental issue by discussing the existence
of a secret tunnel through the black hole's horizon \cite{key-2}.
The energy conservation implies that the black hole contracts during
the process of radiation \cite{key-2,key-3}. Thus, the horizon recedes
from its original radius to a new, smaller radius \cite{key-2,key-3}.
The consequence is that black holes cannot strictly emit thermally
\cite{key-2,key-3}. This is consistent with unitarity \cite{key-2}
and has profound implications for the black hole information puzzle
because arguments that information is lost during black hole's evaporation
rely in part on the assumption of strict thermal behavior of the spectrum
\cite{key-2,key-3}.

\noindent Working with $G=c=k_{B}=\hbar=\frac{1}{4\pi\epsilon_{0}}=1$
(Planck units), the probability of emission is \cite{key-1,key-2,key-3}
\begin{equation}
\Gamma\sim\exp(-\frac{\omega}{T_{H}}),\label{eq: hawking probability}\end{equation}

\noindent where $T_{H}\equiv\frac{1}{8\pi M}$ is the Hawking temperature
and $\omega$ the energy-frequency of the emitted radiation.

\noindent Parikh and Wilczek released a remarkable correction, due
to an exact calculation of the action for a tunneling spherically
symmetric particle, which yields \cite{key-2,key-3}

\noindent \begin{equation}
\Gamma\sim\exp[-\frac{\omega}{T_{H}}(1-\frac{\omega}{2M})].\label{eq: Parikh Correction}\end{equation}

\noindent This important result, which takes into account the conservation
of energy, enables a correction, the additional term $\frac{\omega}{2M}$
\cite{key-2,key-3}. 

\noindent In various frameworks of physics and astrophysics the deviation
from the thermal spectrum of an emitting body is taken into account
by introducing an \emph{effective temperature }which represents the
temperature of a black body that would emit the same total amount
of radiation \cite{key-4}. The effective temperature can be introduced
for black holes too \cite{key-4}. It depends from the energy-frequency
of the emitted radiation and is defined as \cite{key-4}

\noindent \begin{equation}
T_{E}(\omega)\equiv\frac{2M}{2M-\omega}T_{H}=\frac{1}{4\pi(2M-\omega)}.\label{eq: Corda Temperature}\end{equation}

\noindent Then, eq. (\ref{eq: Parikh Correction}) can be rewritten
in Boltzmann-like form \cite{key-4}

\noindent \begin{equation}
\Gamma\sim\exp[-\beta_{E}(\omega)\omega]=\exp(-\frac{\omega}{T_{E}(\omega)}),\label{eq: Corda Probability}\end{equation}

\noindent where $\beta_{E}(\omega)\equiv\frac{1}{T_{E}(\omega)}$
and $\exp[-\beta_{E}(\omega)\omega]$ is the \emph{effective Boltzmann
factor} appropriate for an object with inverse effective temperature
$T_{E}(\omega)$ \cite{key-4}. The ratio $\frac{T_{E}(\omega)}{T_{H}}=\frac{2M}{2M-\omega}$
represents the deviation of the radiation spectrum of a black hole
from the strictly thermal feature \cite{key-4}. If $M$ is the initial
mass of the black hole \emph{before} the emission, and $M-\omega$
is the final mass of the hole \emph{after} the emission \cite{key-3,key-4},
eqs. (\ref{eq: Parikh Correction}) and (\ref{eq: Corda Temperature})
enable the introduction of the \emph{effective mass }and of the \emph{effective
horizon} \cite{key-4} \begin{equation}
M_{E}\equiv M-\frac{\omega}{2},\mbox{ }r_{E}\equiv2M_{E}\label{eq: effective quantities}\end{equation}

\noindent of the black hole \emph{during} the emission of the particle,
i.e. \emph{during} the contraction's phase of the black hole \cite{key-4}.
The \emph{effective quantities $T_{E},$ $M_{E}$ }and\emph{ $r_{E}$
}are average quantities. \emph{$M_{E}$ }is the average of the initial
and final masses, \emph{$r_{E}$ }is the average of the initial and
final horizons and \emph{$T_{E}$ }is the inverse of the average value
of the inverses of the initial and final Hawking temperatures (\emph{before}
the emission $T_{H\mbox{ initial}}=\frac{1}{8\pi M}$, \emph{after}
the emission $T_{H\mbox{ final}}=\frac{1}{8\pi(M-\omega)}$). Notice
that the analyzed process is \emph{discrete} rather than \emph{continuous}.
In fact, the black hole's state before the emission of the particle
and the black hole's state after the emission of the particle are
different countable black hole's physical states separated by an \emph{effective
state} which is characterized by the effective quantities. Hence,
the emission of the particle can be interpreted like a \emph{quantum}
\emph{transition} of frequency $\omega$ between the two discrete
states. The tunneling's visualization is that whenever a tunneling
event works, two separated classical turning points are joined by
a trajectory in imaginary or complex time \cite{key-2}. 

\noindent In this Essay we show that the correction to the thermal
spectrum is also very important for the physical interpretation of
black hole's quasinormal modes \cite{key-4} which in turn results
very important to realize unitary quantum gravity theory as black
holes are considered theoretical laboratories for developing such
an ultimate theory and their quasinormal modes are natural candidates
for an interpretation in terms of quantum levels \cite{key-4,key-5}. 

\noindent The intriguing idea that black hole's quasinormal modes
carry important information about black hole's area quantization is
due to Hod \cite{key-6,key-7}. Hod's original proposal found various
objections over the years \cite{key-5,key-8} which have been answered
in a good way by Maggiore \cite{key-5}, who refined Hod's conjecture.
Quasinormal modes are also believed to probe the small scale structure
of the spacetime \cite{key-9}. 

\noindent The quasinormal frequencies are usually labelled as $\omega_{nl},$
where $l$ is the angular momentum quantum number \cite{key-4,key-5,key-10}.
For each $l$ ($l$$\geq2$ for gravitational perturbations), there
is a countable sequence of quasinormal modes, labelled by the {}``overtone''
number $n$ ($n=1,2,...$) \cite{key-4,key-5}. For large $n$ the
quasinormal frequencies of the Schwarzschild black hole become independent
of $l$ having the structure \cite{key-4,key-5,key-10}

\noindent \begin{equation}
\begin{array}{c}
\omega_{n}=\ln3\times T_{H}+2\pi i(n+\frac{1}{2})\times T_{H}+\mathcal{O}(n^{-\frac{1}{2}})=\\
\\=\frac{\ln3}{8\pi M}+\frac{2\pi i}{8\pi M}(n+\frac{1}{2})+\mathcal{O}(n^{-\frac{1}{2}}).\end{array}\label{eq: quasinormal modes}\end{equation}

\noindent This result was originally obtained numerically in \cite{key-11,key-12},
while an analytic proof was given later in \cite{key-13,key-14}. 

\noindent A problem concerning attempts to associate quasinormal modes
to Hawking radiation was that ideas on the continuous character of
Hawking radiation did not agree with attempts to interpret the frequency
of the quasinormal modes \cite{key-13}. In fact, the discrete character
of the energy spectrum (\ref{eq: quasinormal modes}) should be incompatible
with the spectrum of Hawking radiation whose energies are of the same
order but continuous \cite{key-13}. Actually, the issue that Hawking
radiation is not strictly thermal and, as we have shown, it has discrete
rather than continuous character, removes the above difficulty. In
other words, the discrete character of Hawking radiation permits to
interpret the quasinormal frequencies $\omega_{nl}$ in terms of energies
of physical Hawking quanta too, as we implicitly assumed in \cite{key-4}.
In fact, quasinormal modes are damped oscillations representing the
reaction of a black hole to small, discrete perturbations \cite{key-4,key-5,key-6,key-7}.
A discrete perturbation can be the capture of a particle which causes
an increase in the horizon area \cite{key-5,key-6,key-7}. Hence,
if the emission of a particle which causes a decrease in the horizon
area is a discrete rather than continuous process, it is quite natural
to assume that it is also a perturbation which generates a reaction
in terms of countable quasinormal modes. This natural correspondence
between Hawking radiation and black hole's quasinormal modes permits
to consider quasinormal modes not only for absorbed energies like
in \cite{key-5,key-6,key-7}, but also for emitted energies like in
\cite{key-4}. This issue endorses the idea that, in an underlying
unitary quantum gravity theory, black holes can be considered highly
excited states and their quasinormal modes are the best candidates
for an interpretation in terms of quantum levels \cite{key-4,key-5}. 

\noindent The introduction of the effective temperature $T_{E}(\omega)$
can be applied to the analysis of the spectrum of black hole's quasinormal
modes \cite{key-4}. Another key point is that eq. (\ref{eq: quasinormal modes})
is an approximation as it has been derived with the assumption that
the black hole's radiation spectrum is strictly thermal \cite{key-4}.
To take into due account the deviation from the thermal spectrum in
eq. (\ref{eq: Parikh Correction}) one has to substitute the Hawking
temperature $T_{H}$ with the effective temperature $T_{E}$ in eq.
(\ref{eq: quasinormal modes}) \cite{key-4}. Therefore, the correct
expression for the quasinormal frequencies of the Schwarzschild black
hole, which takes into account the non-strictly thermal behavior of
the radiation spectrum is \cite{key-4}

\noindent \begin{equation}
\begin{array}{c}
\omega_{n}=\ln3\times T_{E}(\omega_{n})+2\pi i(n+\frac{1}{2})\times T_{E}(\omega_{n})+\mathcal{O}(n^{-\frac{1}{2}})=\\
\\=\frac{\ln3}{4\pi(2M-\omega_{n})}+\frac{2\pi i}{4\pi(2M-\omega_{n})}(n+\frac{1}{2})+\mathcal{O}(n^{-\frac{1}{2}}).\end{array}\label{eq: quasinormal modes corrected}\end{equation}

\noindent This important point can be explained as follows \cite{key-4}.
Quasinormal modes are frequencies of the radial spin-j perturbations
$\phi$ of the four-dimensional Schwarzschild background which are
governed by the following master differential equation \cite{key-13,key-14}

\noindent \begin{equation}
\left(-\frac{\partial^{2}}{\partial x^{2}}+V(x)-\omega^{2}\right)\phi.\label{eq: diff.}\end{equation}

\noindent By introducing the Regge-Wheeler potential ($j=2$ for gravitational
perturbations) eq. (\ref{eq: diff.}) is treated as a Schrodinger
equation \cite{key-13,key-14}

\noindent \begin{equation}
V(x)=V\left[x(r)\right]=\left(1-\frac{2M}{r}\right)\left(\frac{l(l+1)}{r^{2}}-\frac{6M}{r^{3}}\right).\label{eq: Regge-Wheeler}\end{equation}

\noindent The relation between the Regge-Wheeler {}``tortoise''
coordinate $x$ and the radial coordinate $r$ is \cite{key-13,key-14}

\noindent \begin{equation}
\begin{array}{c}
x=r+2M\ln\left(\frac{r}{2M}-1\right)\\
\\\frac{\partial}{\partial x}=\left(1-\frac{2M}{r}\right)\frac{\partial}{\partial r}.\end{array}\label{eq: tortoise}\end{equation}

\noindent In \cite{key-13}, Motl derived eq. (\ref{eq: quasinormal modes})
with a rigorous analytical calculation which starts from eqs. (\ref{eq: diff.})
and (\ref{eq: Regge-Wheeler}) and satisfies purely outgoing boundary
conditions both at the horizon ($r=2M$) and in the asymptotic region
($r=\infty$). In order to take into due account the conservation
of energy, one has to substitute the original black hole's mass $M$
in eqs. (\ref{eq: diff.}) and (\ref{eq: Regge-Wheeler}) with the
effective mass of the contracting black hole defined in eq. (\ref{eq: effective quantities})
\cite{key-4}. 

\noindent Hence, eqs. (\ref{eq: Regge-Wheeler}) and (\ref{eq: tortoise})
are replaced by the \emph{effective equations }\cite{key-4}

\noindent \begin{equation}
V(x)=V\left[x(r)\right]=\left(1-\frac{2M_{E}}{r}\right)\left(\frac{l(l+1)}{r^{2}}-\frac{6M_{E}}{r^{3}}\right)\label{eq: effettiva 1}\end{equation}

\noindent and \begin{equation}
\begin{array}{c}
x=r+2M_{E}\ln\left(\frac{r}{2M_{E}}-1\right)\\
\\\frac{\partial}{\partial x}=\left(1-\frac{2M_{E}}{r}\right)\frac{\partial}{\partial r}.\end{array}\label{eq: effettiva 2}\end{equation}

\noindent By realizing step by step the same rigorous analytical calculation
in \cite{key-13}, but starting from eqs. (\ref{eq: diff.}) and (\ref{eq: effettiva 1})
and satisfying purely outgoing boundary conditions both at the effective
horizon ($r_{E}=2M_{E}$) and in the asymptotic region ($r=\infty$),
the final result will be, obviously and rigorously, eq. (\ref{eq: quasinormal modes corrected})
\cite{key-4}.

\noindent An intuitive, elegant interpretation is the following \cite{key-4}.
The imaginary part of (\ref{eq: quasinormal modes}) can be easily
understood \cite{key-14}. The quasinormal frequencies determine the
position of poles of a Green's function on the given background, and
the Euclidean black hole solution converges to a thermal circle at
infinity with the inverse temperature $\beta_{H}=\frac{1}{T_{H}}$
\cite{key-14}. Thus, the spacing of the poles in eq. (\ref{eq: quasinormal modes})
coincides with the spacing $2\pi iT_{H}$ expected for a thermal Green's
function \cite{key-14}. But, if one considers the deviation from
the thermal spectrum it is natural to assume that the Euclidean black
hole solution converges to a \emph{non-thermal} circle at infinity
\cite{key-4}. Therefore, it is straightforward the replacement \cite{key-4}

\noindent \begin{equation}
\beta_{H}=\frac{1}{T_{H}}\rightarrow\beta_{E}(\omega)=\frac{1}{T_{E}(\omega)},\label{eq: sostituiamo}\end{equation}

\noindent which takes into account the deviation of the radiation
spectrum of a black hole from the strictly thermal feature. In this
way, the spacing of the poles in eq. (\ref{eq: quasinormal modes corrected})
coincides with the spacing \cite{key-4}

\noindent \begin{equation}
2\pi iT_{E}(\omega)=2\pi iT_{H}(\frac{2M}{2M-\omega}),\label{eq: spacing}\end{equation}

\noindent expected for a \emph{non-thermal} Green's function (a dependence
on the frequency is present) \cite{key-4}.

\noindent The spectrum of black hole's quasinormal modes can be analysed
in terms of superposition of damped oscillations, of the form \cite{key-4,key-5}
\begin{equation}
\exp(-i\omega_{I}t)[a\sin\omega_{R}t+b\cos\omega_{R}t]\label{eq: damped oscillations}\end{equation}

\noindent with a spectrum of complex frequencies $\omega=\omega_{R}+i\omega_{I}.$
A damped harmonic oscillator $\mu(t)$ is governed by the equation
\cite{key-4,key-5} \begin{equation}
\ddot{\mu}+K\dot{\mu}+\omega_{0}^{2}\mu=F(t),\label{eq: oscillatore}\end{equation}

\noindent where $K$ is the damping constant, $\omega_{0}$ the proper
frequency of the harmonic oscillator, and $F(t)$ an external force
per unit mass. If $F(t)\sim\delta(t),$ i.e. considering the response
to a Dirac delta function, the result for $\mu(t)$ is a superposition
of a term oscillating as $\exp(i\omega t)$ and of a term oscillating
as $\exp(-i\omega t)$, see \cite{key-5} for details. Then, the behavior
(\ref{eq: damped oscillations}) is reproduced by a damped harmonic
oscillator, through the identifications \cite{key-4,key-5}

\noindent \begin{equation}
\begin{array}{ccc}
\frac{K}{2}=\omega_{I}, &  & \sqrt{\omega_{0}^{2}-\frac{K}{4}^{2}}=\omega_{R},\end{array}\label{eq: identificazioni}\end{equation}

\noindent which gives \begin{equation}
\omega_{0}=\sqrt{\omega_{R}^{2}+\omega_{I}^{2}}.\label{eq: omega 0}\end{equation}

\noindent In \cite{key-5} it has been emphasized that the identification
$\omega_{0}=\omega_{R}$ is correct only in the approximation $\frac{K}{2}\ll\omega_{0},$
i.e. only for very long-lived modes. For a lot of black hole's quasinormal
modes, for example for highly excited modes, the opposite limit can
be correct. Maggiore \cite{key-5} used this observation to re-examine
some aspects of quantum physics of black holes that were discussed
in previous literature assuming that the relevant frequencies were
$(\omega_{R})_{n}$ rather than $(\omega_{0})_{n}$. Actually, the
analysis can be further improved by taking into account the important
issue that the radiation spectrum is not strictly thermal \cite{key-4}.
By using the new expression (\ref{eq: quasinormal modes corrected})
for the frequencies of quasinormal modes, one defines \cite{key-4}
\begin{equation}
\begin{array}{ccc}
m_{0}\equiv\frac{\ln3}{4\pi[2M-(\omega_{0})_{n}]}, &  & p_{n}\equiv\frac{2\pi}{4\pi[2M-(\omega_{0})_{n}]}(n+\frac{1}{2}).\end{array}\label{eq: definizioni}\end{equation}

\noindent Then, eq. (\ref{eq: omega 0}) is rewritten in the enlightening
form \cite{key-4} \begin{equation}
(\omega_{0})_{n}=\sqrt{m_{0}^{2}+p_{n}^{2}}.\label{eq: enlightening}\end{equation}

\noindent These results improve eqs. (8) and (9) in \cite{key-5}
as the new expression (\ref{eq: quasinormal modes corrected}) for
the frequencies of quasinormal modes takes into account that the radiation
spectrum is not strictly thermal. For highly excited modes one gets
\cite{key-4}

\noindent \begin{equation}
(\omega_{0})_{n}\approx p_{n}=\frac{2\pi}{4\pi[2M-(\omega_{0})_{n}]}(n+\frac{1}{2}).\label{eq: non equalmente spaziati}\end{equation}
 Thus, differently from \cite{key-5}, levels are \emph{not} equally
spaced even for highly excited modes \cite{key-4}. Indeed, there
are deviations due to the non-strictly thermal behavior of the spectrum
(black hole's effective temperature depends on the energy level). 

\noindent Using eq. (\ref{eq: definizioni}),  one can re-write eq.
(\ref{eq: enlightening}) as \cite{key-4}

\noindent \begin{equation}
(\omega_{0})_{n}=\frac{1}{4\pi[2M-(\omega_{0})_{n}]}\sqrt{(\ln3)^{2}+4\pi^{2}(n+\frac{1}{2})^{2}},\label{eq: enlightening 2}\end{equation}

\noindent which is easily solved giving \cite{key-4} \begin{equation}
(\omega_{0})_{n}=M\pm\sqrt{M^{2}-\frac{1}{4\pi}\sqrt{(\ln3)^{2}+4\pi^{2}(n+\frac{1}{2})^{2}}}.\label{eq: radici}\end{equation}

\noindent As a black hole cannot emit more energy than its total mass,
the physical solution is the one obeying $(\omega_{0})_{n}<M$ \cite{key-4}
\begin{equation}
(\omega_{0})_{n}=M-\sqrt{M^{2}-\frac{1}{4\pi}\sqrt{(\ln3)^{2}+4\pi^{2}(n+\frac{1}{2})^{2}}}.\label{eq: radice fisica}\end{equation}

\noindent The interpretation is of a particle quantized with anti-periodic
boundary conditions on a circle of length \cite{key-4} \begin{equation}
L=\frac{1}{T_{E}(\omega_{0})_{n}}=4\pi\left(M+\sqrt{M^{2}-\frac{1}{4\pi}\sqrt{(\ln3)^{2}+4\pi^{2}(n+\frac{1}{2})^{2}}}\right),\label{eq: lunghezza cerchio}\end{equation}
 i.e. the length of the circle depends on the overtone number $n.$
Maggiore \cite{key-5} found a particle quantized with anti-periodic
boundary conditions on a circle of length $L=8\pi M.$ Our correction
takes into account the conservation of energy, i.e. the additional
term $\frac{\omega}{2M}$ in Eq. (\ref{eq: Parikh Correction}) \cite{key-4}. 

As $(\omega_{0})_{n}$ has to be a real number (an emitted energy),
we need also

\begin{equation}
M^{2}-\frac{1}{4\pi}\sqrt{(\ln3)^{2}+4\pi^{2}(n+\frac{1}{2})^{2}}\geq0\label{eq: need}\end{equation}

in eq. (\ref{eq: radice fisica}). The expression (\ref{eq: need})
is solved giving a maximum value for the overtone number $n$

\begin{equation}
n\leq n_{max}=2\pi^{2}\left(\sqrt{16M^{4}-(\frac{\ln3}{\pi})^{2}}-1\right),\label{eq: n max}\end{equation}

corresponding to $(\omega_{0})_{n_{max}}=M.$ Again, a black hole
cannot emit more energy than its total mass. Thus, the countable sequence
of quasinormal modes for emitted energies is not infinity although
$n$ can be very large. By restoring ordinary units in eq. (\ref{eq: n max}),
one gets, for example, $n_{max}\sim10^{79}$ for a a black hole's
mass of order 10 solar masses.

\noindent Various important consequences on the quantum physics of
black holes arise from the above approach \cite{key-4}. Let us start
with the \emph{area quantization}. 

\noindent Bekenstein \cite{key-15} showed that the area quantum of
the Schwarzschild black hole is $\triangle A=8\pi$ (notice that the
\emph{Planck length} $l_{p}=1.616\times10^{-33}\mbox{ }cm$ is equal
to one in Planck units). By using properties of the spectrum of Schwarzschild
black hole's quasinormal modes, Hod found a different numerical coefficient
\cite{key-6,key-7}. Hod's analysis started by the observation that,
as for the Schwarzschild black hole the \emph{horizon area} $A$ is
related to the mass through the relation $A=16\pi M^{2},$ a variation
$\triangle M$ in the mass generates a variation

\noindent \begin{equation}
\triangle A=32\pi M\triangle M\label{eq: variazione area}\end{equation}

\noindent in the area. By considering an absorption which generates
a transition from an unexcited black hole to a black hole with very
large $n$, Hod assumed \emph{Bohr's correspondence principle} to
be valid for large $n$ and enabled a semi-classical description even
in absence of a full unitary quantum gravity theory \cite{key-6,key-7}.
Thus, from eq. (\ref{eq: quasinormal modes}), the minimum quantum
which can be involved in the transition is $\triangle M=\omega=\frac{\ln3}{8\pi M}.$
This gives $\triangle A=4\ln3.$ The presence of the numerical factor
$4\ln3$ stimulated possible connections with loop quantum gravity
\cite{key-16}. 

\noindent An important criticism by Maggiore \cite{key-5} on Hod's
conjecture is that only transitions from the ground state (i.e. a
black hole which is not excited) to a state with large $n$ (or vice
versa) have been considered by Hod. Bohr's correspondence principle
strictly holds only for transitions from $n$ to $n'$ where both
$n,n'\gg1$ \cite{key-5} and it is also legitimate to consider such
transitions \cite{key-5}. Thus, Maggiore suggested that $(\omega_{0})_{n}$
should be used rather than $(\omega_{R})_{n}$ \cite{key-5}, re-obtaining
the original Bekenstein's result, i.e. $\triangle A=8\pi$. In any
case, Maggiore's result can be also improved if one takes into account
the deviation from the strictly thermal feature in eq. (\ref{eq: Parikh Correction}),
i.e. by using eq. (\ref{eq: quasinormal modes corrected}) rather
than eq. (\ref{eq: quasinormal modes}) \cite{key-4}. From eq. (\ref{eq: radice fisica})
one sees that an emission involving $n$ and $n-1$  gives a variation
of energy \begin{equation}
\triangle M=(\omega_{0})_{n-1}-(\omega_{0})_{n}=-f(M,n)\label{eq: variazione}\end{equation}

\noindent where we have defined \cite{key-4}

\noindent \begin{equation}
\begin{array}{c}
f(M,n)\equiv\\
\\\equiv\sqrt{M^{2}-\frac{1}{4\pi}\sqrt{(\ln3)^{2}+4\pi^{2}(n-\frac{1}{2})^{2}}}-\sqrt{M^{2}-\frac{1}{4\pi}\sqrt{(\ln3)^{2}+4\pi^{2}(n+\frac{1}{2})^{2}}}.\end{array}\label{eq: f(M,n)}\end{equation}

\noindent The sign in (\ref{eq: variazione}) is different, i.e. negative,
with respect to the correspondent eq. (30) in \cite{key-4} because
here we consider an emission while in \cite{key-4} we considered
an absorption.

\noindent Combining eqs. (\ref{eq: variazione area}) and (\ref{eq: variazione})
one gets \cite{key-4}

\noindent \begin{equation}
\triangle A=32\pi M\triangle M=-32\pi M\times f(M,n).\label{eq: area quantum}\end{equation}

\noindent For very large $n$ (but we recall that $n\leq n_{max}$,
see eq. (\ref{eq: n max})) one obtains \cite{key-4}

\noindent \begin{equation}
\begin{array}{c}
f(M,n)\approx\\
\\\approx\sqrt{M^{2}-\frac{1}{2}(n-\frac{1}{2})}-\sqrt{M^{2}-\frac{1}{2}(n+\frac{1}{2})}\approx\frac{1}{4M},\end{array}\label{eq: circa}\end{equation}

\noindent and eq. (\ref{eq: area quantum}) becomes $\triangle A\approx-8\pi$
which is the original result of Bekenstein for the area quantization
(a part a sign because we consider an emission rather than an absorption).
Then, only in the very large $n$ limit the levels are approximately
equally spaced \cite{key-4}. Indeed, for smaller $n$ there are deviations,
see eq. (\ref{eq: non equalmente spaziati}).

\noindent Important consequences on entropy and microstates arise
from the above analysis \cite{key-4}.

\noindent Let us assume that, for large $n$, the horizon area is
quantized \cite{key-5} with a quantum $|\triangle A|=\alpha,$ where
$\alpha=32\pi M\cdot f(M,n)$ for us \cite{key-4}, $\alpha=8\pi$
for Bekenstein \cite{key-15} and Maggiore \cite{key-5}, $\alpha=4\ln3$
for Hod \cite{key-6,key-7}. The total horizon area must be $A=N|\triangle A|=N\alpha$
(notice that the number of quanta of area, the integer $N,$ is \emph{not}
the overtone number $n$). Our approach gives \cite{key-4}

\noindent \begin{equation}
N=\frac{A}{|\triangle A|}=\frac{16\pi M^{2}}{\alpha}=\frac{16\pi M^{2}}{32\pi M\cdot f(M,n)}=\frac{M}{2f(M,n)}.\label{eq: N}\end{equation}
 The famous formula of Bekenstein-Hawking entropy \cite{key-1,key-17,key-18}
now becomes \cite{key-4}

\noindent \begin{equation}
S_{BH}=\frac{A}{4}=8\pi NM|\triangle M|=8\pi NM\cdot f(M,n).\label{eq: Bekenstein-Hawking}\end{equation}

\noindent Thus, we get the important result that Bekenstein-Hawking
entropy is a function of the quantum overtone number $n$.

\noindent In the very large $n$ limit eq. (\ref{eq: circa}) gives
$f(M,n)\rightarrow\frac{1}{4M}$ and the standard result \cite{key-5,key-19,key-20,key-21}

\noindent \begin{equation}
S_{BH}\rightarrow2\pi N\label{eq: entropia circa}\end{equation}

\noindent is re-obtained \cite{key-4}.

\noindent On the other hand, it is a common and general belief that
there is no reason to expect that Bekenstein-Hawking entropy will
be the whole answer for a correct unitary quantum gravity theory \cite{key-22}.
For a better understanding of black hole's entropy one needs to go
beyond Bekenstein-Hawking entropy and identify the sub-leading corrections
\cite{key-22}. The quantum tunnelling approach can be used to obtain
the sub-leading corrections to the second order approximation \cite{key-23}.
One gets that the black hole's entropy contains three parts: the usual
Bekenstein-Hawking entropy, the logarithmic term and the inverse area
term \cite{key-23} 

\noindent \begin{equation}
S_{total}=S_{BH}-\ln S_{BH}+\frac{3}{2A}.\label{eq: entropia totale}\end{equation}

\noindent In fact, if one wants to satisfy the unitary quantum gravity
theory the logarithmic and inverse area terms are requested \cite{key-23}.
Apart from a coefficient, this correction to the black hole's entropy
is consistent with the one of loop quantum gravity \cite{key-23},
where the coefficient of the logarithmic term has been rigorously
fixed at $\frac{1}{2}$ \cite{key-23,key-24}. The correction (\ref{eq: Bekenstein-Hawking})
to Bekenstein-Hawking entropy permits to re-write eq. (\ref{eq: entropia totale})
as \cite{key-4}

\noindent \begin{equation}
S_{total}=8\pi NM\cdot f(M,n)-\ln8\pi NM\cdot f(M,n)+\frac{3}{64\pi NM\cdot f(M,n)}\label{eq: entropia totale 2}\end{equation}

\noindent that in the very large $n$ limit becomes \cite{key-4}

\noindent \begin{equation}
S_{total}\rightarrow2\pi N-\ln2\pi N+\frac{3}{16\pi N}.\label{eq: entropia totale approssimata}\end{equation}

\noindent These results imply that at level $N$ the black hole has
a number of microstates \cite{key-4}

\noindent \begin{equation}
g(N)\propto\exp\left[8\pi NM\cdot f(M,n)-\ln8\pi NM\cdot f(M,n)+\frac{3}{64\pi NM\cdot f(M,n)}\right],\label{eq: microstati}\end{equation}

\noindent that in the very large $n$ limit reads \cite{key-4}

\noindent \begin{equation}
g(N)\propto\exp(2\pi N-\ln2\pi N+\frac{3}{16\pi N}).\label{eq: microstati circa}\end{equation}

\noindent In summary, in this Essay the discrete character of Hawking
radiation, which is due to the non-strict thermal behavior of the
spectrum, has been used to enable a natural correspondence between
Hawking radiation and black hole's quasinormal modes. This important
issue endorses the idea that, in an underlying unitary quantum gravity
theory, black holes result highly excited states. This key point permits
to re-analyze the spectrum of black hole's quasinormal modes by introducing
the black hole's effective temperature (\ref{eq: Corda Temperature}).
The analysis enables a re-examination of various results in the literature
by changing the physical understanding of the radiation spectrum.
In this way, important modifies on quantum physics of black holes
have been realized. In particular, the formula of the horizon's area
quantization and the number of quanta of area have been modified becoming
functions of the quantum overtone number $n$. Hence, Bekenstein-Hawking
entropy, its sub-leading corrections and the number of microstates,
which are fundamental to realize unitary quantum gravity theory, have
also been modified. These quantities become functions of the quantum
overtone number too. 

\noindent Previous results in the literature are re-obtained in the
very large $n$ limit. This point confirms the correctness of the
analysis in this Essay which improves previous approximations.

\subsubsection*{Acknowledgements}

It is a pleasure to thank Alexander Burinskii, Lawrence B. Crowell,
Herman J. Mosquera Cuesta, Jeremy Dunning-Davies, Boris P. Kosyakov,
Jorge Ovalle, Erasmo Recami, Izzet Sakalli, Douglas Singleton, and
George Tsoupros for various interesting comments and discussions on
black holes' physics.

\end{document}